\documentclass[aps,prd,twocolumn,amsmath,superscriptaddress,amssymb,showpacs,floatfix,nofootinbib,longbibliography,preprintnumbers]{revtex4-1}
\pdfoutput=1
\usepackage{graphicx}
\usepackage{bm}
\usepackage{times}
\usepackage{slashed}
\usepackage{color}
\usepackage{slashed}
\usepackage{lipsum}
\usepackage{subfigure}
\usepackage{multirow}
\usepackage{amsmath}
\usepackage{array} 
\usepackage{varwidth} 
\usepackage[T1]{fontenc}
\usepackage[utf8]{inputenc}
\usepackage[english]{babel}
\usepackage{amssymb}
\usepackage{amsmath}
\usepackage{textgreek}

\newcommand{\be}{\begin{equation}}
\newcommand{\ee}{\end{equation}}
\newcommand{\bea}{\begin{eqnarray}}
\newcommand{\eea}{\end{eqnarray}}

\def\lsim{\mathrel{\raise.3ex\hbox{$<$\kern-.75em\lower1ex\hbox{$\sim$}}}}
\def\gsim{\mathrel{\raise.3ex\hbox{$>$\kern-.75em\lower1ex\hbox{$\sim$}}}}

\begin{document}

\title{The Contribution From TeV Halos to the Isotropic Gamma-Ray Background}
\author{Fei Xu}
\email{feixu@uchicago.edu, ORCID: orcid.org/0000-0002-8850-7580}
\affiliation{Department of Astronomy and Astrophysics, University of Chicago, 5640 South Ellis Ave., Chicago, IL 60637}

\author{Dan Hooper}
\email{dhooper@fnal.gov, ORCID: orcid.org/0000-0001-8837-4127}
\affiliation{Theoretical Astrophysics Department, Fermi National Accelerator Laboratory, Batavia, Illinois, 60510, USA}
\affiliation{Department of Astronomy and Astrophysics, University of Chicago, 5640 South Ellis Ave., Chicago, IL 60637}
\affiliation{Kavli Institute for Cosmological Physics, The University of Chicago, Chicago, IL 60637, USA}

\begin{abstract}
Recent observations have shown that pulsars are surrounded by extended regions which emit TeV-scale gamma rays through the inverse Compton scattering of very high energy electrons and positrons. Such TeV halos are responsible for a large fraction of the Milky Way's TeV-scale gamma-ray emission. In this paper, we calculate the gamma-ray spectrum from the population of TeV halos located within the Andromeda Galaxy, predicting a signal that is expected to be detectable by the Cherenkov Telescope Array (CTA). We also calculate the contribution from TeV halos to the isotropic gamma-ray background (IGRB), finding that these sources should contribute significantly to this flux at the highest measured energies, constituting up to $\sim 20\%$ of the signal observed above $\sim 0.1 \, {\rm TeV}$.
\end{abstract}

\maketitle

\section{Introduction}
Observations with the HAWC (High Altitude Water Cherenkov)~\cite{Hooper:2017gtd,Linden:2017vvb,HAWC:2017kbo,HAWC:2019tcx,HAWC:2020hrt}, Milagro~\cite{2009ApJ...700L.127A, Linden:2017blp}, and HESS (High Energy Spectroscopic System)~\cite{HESS:2018pbp,HESS:2017lee} telescopes have revealed that pulsars are surrounded by spatially extended ``TeV halos''~\cite{HAWC:2021dtl}. The multi-TeV gamma-ray emission that is associated with these halos is the result of inverse Compton scattering, and is powered by the rotational kinetic energy of the host pulsar~\cite{sudoh2021highest}. These objects represent a new class of high-energy sources, which are responsible for a significant fraction of the Milky Way's TeV-scale gamma-ray emission.

From the measured abundance of pulsars and the efficiency with which they are observed to generate TeV halos, it can be shown that these objects dominate the diffuse TeV-scale emission that is observed along the plane of the Milky Way~\cite{Linden:2017blp}. On similar grounds, one can deduce that this class of sources must contribute significantly to the total isotropic gamma-ray background (IGRB), in particular at TeV-scale energies. In this sense, TeV halos appear to be an important means by which star formation leads to the production of very high-energy radiation.

In this paper, we use the observed characteristics of the TeV halos detected by HAWC to estimate the TeV-scale gamma-ray emission from the TeV halo population in the Andromeda Galaxy (M31), as well as to calculate the contribution from this source class to the total IGRB. While we conclude that the gamma-ray emission from the TeV halos in M31 should be below current constraints, we predict that this signal will be detectable in the future by the Cherenkov Telescope Array (CTA). Furthermore, we predict that the TeV halos distributed throughout the observable Universe contribute significantly to the IGRB, being responsible for up to $\sim$20\% of this background at 100 GeV, and perhaps even a larger fraction at TeV-scale energies. Furthermore, if the total spindown power of the millisecond pulsar population is comparable to or larger than that associated with young and middle aged pulsars, this would significantly increase our estimates for these gamma-ray fluxes.

\section{Gamma-Ray Emission From TeV Halos}

Pulsars generate the gamma-ray emission associated with TeV halos by transferring their rotational kinetic energy into the acceleration of very high-energy electrons and positrons. These particles then diffuse outward and undergo inverse Compton scattering. The integrated energy budget for the resulting emission is, therefore, limited by the pulsar's initial rotational kinetic energy, which in turn depends on its initial period and moment of inertia: 
\begin{equation}
    E_{\rm rot} = \frac{I \Omega^2}{2} = \frac{4\pi^2 M R^2}{5P^2},
\end{equation}
\noindent where $M$ and $R$ are the mass and radius of the neutron star.
By extension, the time-averaged, total energy budget for a population of TeV halos is given by the product of the pulsar birth rate and the average initial rotational kinetic energy of an individual pulsar. With this in mind, we will consider the value of $E_{\rm rot}$ averaged over an ensemble of newly formed pulsars:
\begin{align}
\langle E_{\rm rot, 0} \rangle \approx  \frac{4 \pi^2 M R^2}{5} \bigg\langle \frac{1}{P^2_0} \bigg\rangle   
\end{align}
The pulsar-to-pulsar variations in $M$ and $R$ are each small compared to those associated with a pulsar's initial period. Throughout this study, we adopt $M = 1.28 \, M_{\odot}$ \citep{_zel_2012} and $R = 11.9 \,\rm km$ \citep{Abbott_2018}.
%$M = 1.28\pm 0.24 \m M_{\odot}$ \citep{_zel_2012} and $R = 11.9 \pm 1.4\,\rm km$ \citep{Abbott_2018}. 
The initial period of a pulsar, $P_0$, can be difficult to determine directly from observations~\cite{Kaspi_2001, Popov_2012, Igoshev_2013}. To estimate the value of $\langle P_0^{-2}\rangle$, we have performed an average of this quantity over the youngest pulsars contained within the Australia Telescope National Facility (ATNF) pulsar catalog~\cite{Manchester:2004bp}. The evolution of a pulsar's period is described by
\begin{align}
P(t) = P_0 \bigg(1+\frac{t}{\tau}\bigg)^{1/(n-1)},
\end{align}
where $n$ is the braking index ($n=3$ in the case that the pulsar's spin-down torque arises entirely from dipole radiation~\cite{1969ApJ...157.1395O, Johnston_1999}) and $\tau$ is the spindown timescale:
\begin{align}
\tau &= \frac{3c^2 I P^2_0}{4 \pi^2 B^2 R^6} \\
&\approx 3.5 \times 10^4 \, {\rm yr} \times \bigg(\frac{2 \times 10^{12}\,{\rm G}}{B}\bigg)^2 %\,\bigg(\frac{M}{1.28 \, M_{\odot}}\bigg) \, \bigg(\frac{11.9 \, {\rm km}}{R}\bigg)^4 
\, \bigg(\frac{P_0}{0.065 \, {\rm s}}\bigg)^2. \nonumber
\end{align}
With this timescale for spindown in mind, we performed an average of $P^{-2}$ over the 32 (non-binary) pulsars in the ATNF catalog with a characteristic age of $10^4$ years or less, defined as $t_c \equiv P/2\dot{P}= (n-1)(t_{\rm age}+\tau)/2 < 10^4 \, {\rm yr}$, finding $\langle P^{-2}\rangle = (65 \, {\rm ms})^{-2}$. From this, we estimate that the initial rotational kinetic energy of an average pulsar is $E_{{\rm rot},0} \approx  7 \times 10^{48} \, {\rm erg}$. Note that among this sample, there is no discernible correlation between the pulsars' period and distance, suggesting that no sizable bias is likely to have resulted from selection effects. If we expand our sample to consider the 151 pulsars with $t_c < 10^5 \, {\rm yr}$, we obtain a somewhat lower average rate of rotation, $\langle P^{-2}\rangle = (95 \, {\rm ms})^{-2}$, indicative of a non-negligible reduction in the average pulsar's rotational kinetic energy.

Only a fraction of a given pulsar's total rotational kinetic goes into the gamma-ray emission associated with a TeV halo. We define the efficiency of a TeV halo, $\eta$, as the fraction of the pulsar's rotational kinetic energy that goes into the production of TeV-scale gamma-rays:  
\begin{equation}
\eta \equiv \frac{F_{\gamma}}{\dot{E}_{\rm rot}/4\pi d^2},
\end{equation}
where $F_{\gamma}$ is the flux of the gamma-ray emission bewteen 0.1-100 TeV, $\dot{E}_{\rm rot}$ is the time derivative of the pulsar's rotational kinetic energy, and $d$ is the distance to the pulsar. Once again, we are interested here in the value of $\eta$ averaged across a large sample of pulsars. We determine this quantity by comparing the current spindown flux of a given pulsar, $\dot{E}_{\rm rot}/4\pi d^2$, as reported in the ATNF catalog, to the gamma-ray flux reported by the HAWC Collaboration, as obtained using the tool available at \url{https://data.hawc-observatory.org/datasets/3hwc-survey/index.php}. In making this comparison, we have restricted our sample to those pulsars that are located within HAWC's field-of-view ($-10^{\circ} < {\bf {\rm dec}} < 50^{\circ}$), and for which $t_{c} > 10^4$ years. This latter requirement is intended to avoid contaminating our sample with sources that might be better classified as pulsar wind nebulae or supernova remnants. We also restrict our analysis to those pulsars with a spindown flux greater than $\dot{E}/4\pi d^2 > 10^{-10}  \, {\rm TeV} \, {\rm cm}^{-2} \, {\rm s}^{-1}$, in an effort to minimize any bias that might result from selection effects. We have identified 26 pulsars in the ATNF catalog which satisfy these criteria.

The HAWC online tool allows one to obtain a measurement of the gamma-ray flux from a given source, as evaluated at an energy of 7 TeV, assuming a power-law spectrum with an index of -2.5. For each pulsar, we integrate over this spectral shape between 0.1 and 100 TeV to obtain an estimate for $F_{\gamma}$. Following Refs.~\cite{hooper2021evidence,Hooper:2018fih}, we adopt the point-like template for pulsars located at $d >$ 2 kpc, the template with 0.5$^{\circ}$ extension for pulsars between 0.75 kpc $< d$ $<$ 2 kpc, the 1$^{\circ}$ extension template for those within 0.375 kpc $< d$ $<$ 0.75 kpc, and the 2$^{\circ}$ extension template for pulsars closer than $d$ $<$0.375 kpc. Averaging over this sample, we obtain an average gamma-ray efficiency of $\langle \eta \rangle = 0.054$.

We note that the HAWC online tool is not very flexible in the respect that it only constrains the flux from a given source assuming that its spectrum is described by a power-law with an index of -2.5, and thus is not optimally suited for the application at hand. In particular, while the detailed spectral shape of the gamma-ray emission from a TeV halo has been measured only in a few cases, these sources appear to exhibit spectra that are significantly harder than that of a -2.5 index power law. More specifically, the gamma-ray emission from TeV halos is observed to be quite hard up to energies on the order of $\mathcal{O}(10 \, {\rm TeV})$, above which the spectrum becomes much softer. On theoretical grounds, one expects such a spectral break to appear, positioned near the energy at which the timescale for electron/positron energy losses are comparable to the age of the pulsar~\cite{sudoh2021highest}. In light of these considerations, it is plausible that the harder spectra indices of TeV halos may have led us to somewhat overestimate the value of $\langle \eta \rangle$ in the approach taken in the previous paragraph.

In the 3HWC catalog presented by the HAWC Collaboration, the flux and spectral index of each source is provided, as evaluated at an energy of 7 TeV~\cite{HAWC:2020hrt}. More information, however, is provided for some of these sources in HAWC's catalog of TeV halos detected at energies above 56 TeV~\cite{HAWC:2019tcx}. Averaging the value of $\eta$ over this collection of 9 sources (see Table 1 of Ref.~\cite{sudoh2021highest}), we obtain $\langle \eta \rangle \approx 0.063$, which is only slightly higher than the value found using the approach described in the previous paragraph. In light of these considerations, we will adopt a range of $\langle \eta \rangle = 0.04 - 0.06$ throughout the remainder of this study.

To assess the spectral shape of the gamma-ray emission from a typical TeV halo, we consider three sources which have had their spectra measured in some detail~\cite{HAWC:2019tcx}. In particular, we will base our results on the spectral shapes of the emission observed from eHWC J1825-134 (PSR J1826-1256), eHWC J1907+063 (PSR J1907+0602), and eHWC J2019+368 (PSR J2021+3651), as reported in Ref.~\cite{HAWC:2019tcx} (see also, Ref~\cite{sudoh2021highest}). These sources each exhibit a spectrum that can be reasonably well described by a smoothly broken power-law, which we parameterize as follows:
\begin{equation}
\begin{aligned}
\frac{dN_{\gamma}}{dE_{\gamma}} \propto \bigg(\frac{E_{\gamma}}{E_b}\bigg)^{-\alpha} \bigg[1+\bigg(\frac{E_{\gamma}}{E_b}\bigg)\bigg]^{\alpha-\beta}. 
%\frac{E_{\gamma}^{\alpha}}{1+(\frac{E_{\gamma}}{E_p})^{\alpha-\beta}}
\end{aligned}
\end{equation}

For the three above mentioned TeV halos, the spectrum of inverse Compton scattering given in Fig.~3 of Ref.~\cite{sudoh2021highest} is best fit by ($\alpha$, $\beta$, $E_b$) = $(1.65, 3.36, 5.9\, {\rm TeV})$, $(1.58, 3.08, 6.2 \, {\rm TeV})$, and $(1.66, 3.12, 23.6 \, {\rm TeV})$, respectively. Based on these selected sources, we adopt $\alpha=1.63$, $\beta=3.18$, and $E_b=10 \,{\rm TeV}$ as our estimate for the spectral shape of a typical TeV halo.

\section{Gamma-Ray Emission from Andromeda's TeV Halo Population}
\label{sec:Andromeda}

Before moving forward to calculate the total gamma-ray emission from the TeV halos found throughout the volume of the observable Universe, we will consider in this section the prospects for detecting such a signal from the TeV halos located within the Andromeda Galaxy. The Andromeda galaxy, or M31, is a spiral galaxy located at a distance of $d_{\rm M31} = $765$\pm$28 kpc from the Milky Way~\citep{Riess_2012}. By comparing its current rate of star formation to that of the Milky Way's, we will estimate the total gamma-ray emission from M31's TeV halo population and compare this to the sensitivity of existing and future gamma-ray telescopes. 

To estimate the current pulsar birth rate in M31, $\Gamma_{p,{\rm M31}}$, we assume that this quantity scales with the overall star-formation rate, $\Gamma_{\star}$, and thereby relate the pulsar birth rate in M31 to that in the Milky Way as follows:
\begin{equation}
\Gamma_{p,{\rm M31}} = \frac{\Gamma_{\star,{\rm M31}}}{\Gamma_{\star,{\rm MW}}} \times \Gamma_{p,{\rm MW}}.
\end{equation}
While this relationship is only expected to apply to galaxies which produce stars with a similar initial mass function, it should be safely applicable in the particular case of Andromeda and the Milky Way.

Many methods are used to determine or constrain the star-formation rate of a given galaxy, including those based on observations of Lyman continuum photons, infrared emission, H$\alpha$ lines, ultra-violet emission, supernovae rates, and counts of resolved stellar populations (for reviews, see Refs.~\cite{Kennicutt:2012ea,Madau_2014}). Many of these techniques are sensitive to the rate of massive star formation, which can be extrapolated to determine the total star-formation rate for a given choice of the inital mass function. Some of these techniques can be applied to the case of the Milky Way, while others are more suitable to other galaxies~\cite{Chomiuk_2011}. For example, H$\alpha$ emission is often used to estimate the star-formation rates of galaxies, but is not useful in the plane of the Milky Way due to the effects of dust extinction~\cite{Chomiuk_2011}. In this paper, we adopt for the Milky Way a star-formation rate given by $\Gamma_{\star, {\rm MW}} = 1.65 \pm 0.19 \, M_{\odot} \rm yr^{-1}$, based on a combination of measurements including the Lyman continuum photon flux, supernovae rates, massive star counts, and infrared emission~\cite{Licquia_2015}. For the case of M31, we follow Ref.~\cite{Rahmani_2016}, which describes three methods for measuring the star-formation rate of Andromeda. Using a combination of far-UV and 24\,$\mu m$ emission, H$\alpha$ emission and 24 $\mu m$ emission, and the total infrared emission, that study obtained $\Gamma_{\star, {\rm M31}} = 0.31\pm0.04 \, M_{\odot} \rm yr^{-1}$, $0.35\pm0.01 \, M_{\odot} \rm yr^{-1}$, and $0.40\pm0.04 \, M_{\odot} \rm yr^{-1}$, respectively. With these results in hand, we adopt a range for $\Gamma_{\star, {\rm M31}}/\Gamma_{\star, {\rm MW}}$ that is given by {$0.21 \pm 0.04$.} After combining this in quadrature with a value of $1.4 \pm 0.2$ pulsars per century for the Milky Way's pulsar birth rate~\cite{2006MNRAS.372..777L}, this yields a birth rate of {$0.29 \pm 0.07$} pulsars per century in M31. Alternatively, the observed rates for core-collapse supernovae (CCSN) can be used to estimate the ratio of pulsar birth rates in the Milky Way and Andromeda. Based on Ref.~\cite{Rozwadowska_2021}, this yields ${\rm CCSN}_{\rm M31}/{\rm CCSN}_{\rm MW} \sim 0.25-0.75$, corresponding to a pulsar birth rate of $\sim 0.3-1.2$ per century for M31, consistent with our previous determination.

\begin{figure*}
\centering
\includegraphics[height=8cm,width=11cm]{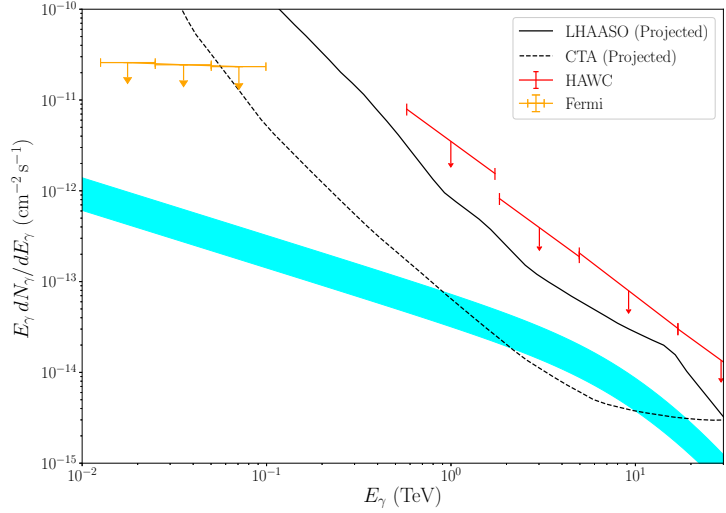}
\caption{The gamma-ray emission from the population of TeV halos in the Andromeda Galaxy (M31) is shown as a cyan band. The width of this band reflects the uncertainties in the gamma-ray efficiency and pulsar birth rate, for which we have adopted the following ranges: $\langle \eta \rangle = 0.04-0.06$ and $\Gamma_{p,{\rm M31}} = 0.23-0.35$ per century. These results are compared to the upper limits reported by the HAWC~\cite{Albert_2020} and Fermi~\cite{Ackermann_2017} Collaborations, as well as the projected sensitivity of LHAASO~\cite{bai2019large} and CTA~\cite{CTAConsortium:2013ofs} (for 1 year and 50 hours of observation, respectively). While our range of estimates for this emission are consistent with current constraints, the prospects for detecting this emission with future telescopes appear promising.}
\label{fig:M31 spec}
\end{figure*}

Using this calculation for the pulsar rate in M31, we can now estimate the total TeV halo emission from this galaxy. We will proceed under the reasonable assumption that the TeV halos in M31 generate gamma-ray emission that is similar in overall intensity and spectral shape to those found in the Milky Way. The total TeV-scale luminosity of the TeV halos in M31 can thus be expressed as $L_{\rm M31} = \Gamma_{p, {\rm M31}} \, \langle \eta \rangle \, \langle E_{{\rm rot}, 0} \rangle/4\pi d^2$. 

In Fig.~\ref{fig:M31 spec}, we plot our estimate for the gamma-ray emission from the TeV halo population of M31. The width of this band reflects the uncertainties in the gamma-ray efficiency and pulsar birth rate. We compare this result to the upper limits reported by the HAWC~\cite{Albert_2020} and Fermi~\cite{Ackermann_2017} (see also, Refs.~\cite{Karwin:2019jpy,Karwin:2020tjw}) Collaborations, as well as the projected sensitivity of LHAASO~\cite{bai2019large} and CTA~\cite{CTAConsortium:2013ofs}. These projected sensitivities were each calculated by simulating the detector response to a Crab Nebula-like point source, and adopting an observation time of 1 year (LHAASO) or 50 hours (CTA). While our projections for this emission are consistent with current constraints, the prospects for detecting this emission with future telescopes seem promising. Note that we expect TeV halos to provide the dominant contribution to the $\gtrsim {\rm TeV}$ gamma-ray emission from galaxies such as the Milky Way and M31~\cite{Linden:2017blp}.

In addition to characterizing the gamma-ray emission originating from TeV halos, future gamma-ray observations of M31 will also provide valuable information pertaining to cosmic ray transport in that system, and will constrain more exotic signals, such as emission from Andromeda supermassive black hole, emission analogous to the Milky Way's ``Fermi Bubbles'', and the products of dark matter annihilation or decay~\cite{2003A&A...400..153A, McDaniel_2019, Ackermann_2017, Albert_2020}.

\section{TeV Halos and the Isotropic Gamma-Ray Background}

In the previous section, we calculated the emission from TeV halos in the nearby galaxy M31. In this section, we will proceed to calculate the total emission from TeV halos throughout the observable Universe, determining their contribution to the isotropic gamma-ray background (IGRB) as measured by the Fermi telescope~\cite{Ackermann_2015}.

Neglecting the effects of attenuation for the moment, the spectrum of gamma rays per area per time per solid angle from the integrated population of extragalactic TeV halos is given by:
\begin{align}
\frac{dN_{\gamma}}{dE_{\gamma}}(E_{\gamma}) &= \frac{c}{4\pi} \int \frac{dz}{H(z)(1+z)^3} \\
& \times \frac{d\Gamma_p}{dV}(z) \langle E_{{\rm rot}, 0} \rangle \, \langle \eta \rangle  \,\bigg(A\frac{dN_{\gamma}}{dE'}\bigg)_{E'=E_{\gamma}(1+z)},    \nonumber
\end{align}
where $H(z) = H_0 \,[\Omega_M(1+z)^3+\Omega_{\Lambda}]^{0.5}$ is the rate of Hubble expansion, $d\Gamma_p/dV (z)$ is the average pulsar birth rate per volume as a function of redshift, the quantity $\langle E_{{\rm rot}, 0} \rangle \times \langle \eta \rangle$ is the average total energy emitted from a pulsar in TeV-scale gamma-rays, and $dN_{\gamma}/dE'$ is the average spectrum of the gamma-ray emission from an individual pulsar, after accounting for the effects of cosmological redshift. The normalization constant, $A$, has units of inverse energy, and is set such that
\begin{align}
\int_{0.1 \, {\rm TeV}}^{100 \, {\rm TeV}} A \frac{dN_{\gamma}}{dE'} E' dE' = 1.
\end{align}
Throughout this study, we will adopt $\Omega_M=0.31$, $\Omega_{\Lambda}=0.69$, and $H_0=67.7 \, {\rm km /s/Mpc}$, as reported by the Planck Collaboration~\cite{Planck:2018vyg}.

\begin{figure*}
\centering
\includegraphics[height=8cm,width=11cm]{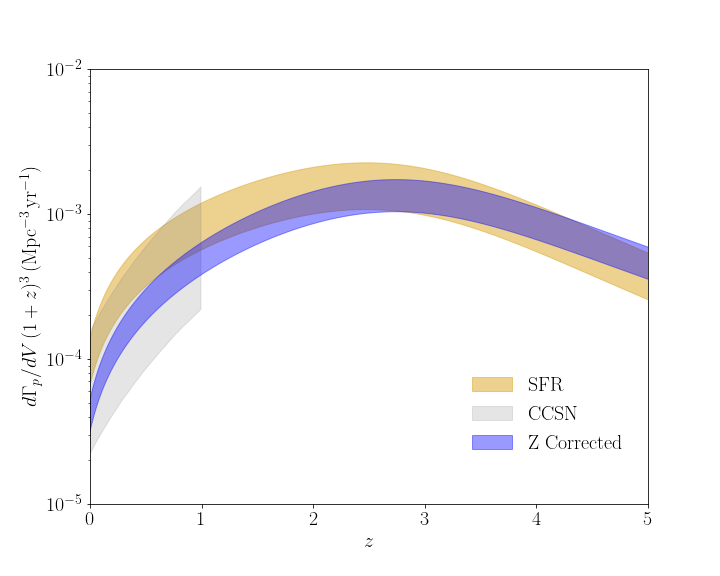}
\caption{The comoving pulsar birth rate density as a function of redshift, calculated based on the star-formation rate density (yellow), the core collapse supernova rate density (grey), and using the metallicity corrected method described in the text (blue). The bands reflect the $1\sigma$ uncertainties in the measurements of the star-formation rate density and the core collapse supernova rate.}
\label{fig:dNdV}
\end{figure*}

As we did in the case of M31, we will base our estimate for the emission from the sum of all cosmologically distributed TeV halos on that observed from these objects in the Milky Way, scaling the relative intensities with the pulsar birth rate. To estimate the pulsar birth rate as a function of redshift, we adopt three different approaches. First, we estimate the pulsar birth rate as a function of redshift by scaling this function to the measured rate of star formation. More specifically, we adopt the cosmic star-formation rate (per comoving volume) as reported in Ref.~\cite{Hopkins_2006} (and using the parametric form of Ref.~\cite{Cole_2001}):
%\Gamma_{p,{\rm M31}}
\begin{equation}
\begin{aligned}
\dot{\rho_{*}}(z) = \frac{(0.017+0.13 z)h}{1+(z/3.3)^{5.3}} \, \rm M_{\odot} \, Mpc^{-3} \, yr^{-1},
\end{aligned}
\end{equation}
where $h=0.677$. The $1 \sigma$ uncertainty associated with this quantity is approximately $\pm 25\%$ \citep{Horiuchi_2011}. Assuming that the pulsar birth rate is proportional to the star-formation rate, we can relate $d\Gamma_p/dV (z)$ to this function, and to the local pulsar birth rate to star-forming rate ratio:
\begin{align}
\frac{d\Gamma_p}{dV}(z) = \frac{\rho_*(z)}{(1+z)^3} \, \frac{\Gamma_{p, {\rm MW}}}{\Gamma_{\star, {\rm MW}}}. 
\end{align}

Alternatively, we could instead scale the pulsar birth rate to the rate of core collapse supernovae as measured, for example, by the Lick Observatory Supernova Search (LOSS) \cite{2011MNRAS.412.1419L, 2011MNRAS.412.1441L, 2011MNRAS.412.1473L, 2011}, and then normalize this to the rate of core collapse supernovae in the Milky Way, $R_{\rm CCSN, MW} = 1.9\pm 1.1$ per century~ \cite{2006Natur.439...45D}. This approach has the advantage of being less sensitive to variations in the initial mass function, but suffers from larger overall uncertainties and is limited to modest redshifts, $z \lesssim 1$.

As a third method, we have estimated the pulsar birth rate over cosmic history from the evolution of the initial mass function as a function of mass and metallicity. From the initial mass function, we can calculate the number of neutron stars that are formed per unit mass of star formation:
\begin{equation}
f_{\rm NS} = \frac{\int_{M_{\rm min}}^{M_{\rm max}} \phi \, dM}{\int_{0.1 M_{\odot}}^{100 M_{\odot}} \phi \, M dM},
\end{equation}
where $M_{\rm min}$ and $M_{\rm max}$ represent the mass range of stellar progenitors that ultimately lead to the formation of a neutron star. The function $\phi$ is the initial mass function, for which we adopt the following~\cite{Marks_2012}:

\begin{center}
 \[ \phi (M, Z) \propto \begin{cases}
M^{-1.3} & 0.1 M_{\odot}<  M <0.5 M_{\odot} \\
M^{0.66 \log_{10} (Z/Z_{\odot})+2.63} & 0.5 M_{\odot}< M< 100 M_{\odot}
\end{cases} \]
\end{center}

\noindent where $Z/Z_{\odot}$ is the metallicity in solar units. Notice that the slope at low masses follows the canonical behavior of the~\citet{2001MNRAS.322..231K} initial mass function. 

Depending on the mass of the final remnant, a core collapse supernova can produce a neutron star or a black hole. We set the threshold for this distinction to $2.5 M_{\odot}$, which we then relate to the maximum initial stellar mass, $M_{\rm max}$, as a function of metallicity according to Eqns.~5-9 in Ref.~\cite{Fryer_2012}. We then determine as follows the minimum initial stellar mass, $M_{\rm min}$, that can result the formation of a neutron star~\cite{Fryer_2012}:

 \[  M_{\rm min}=\begin{cases}
[9.0 + 0.9\,\log_{10}(Z/Z_{\odot})]\, M_{\odot} & \log_{10}(Z/Z_{\odot})>-3 \\
6.3 \,M_{\odot} & \log_{10}(Z/Z_{\odot})\leq -3
\end{cases} \]

Since the quantities $\phi$, $M_{\rm max}$ and $M_{\rm min}$ each depend on metallicity, we need to quantify the distribution of $Z$ as a function of redshift. To this end, we follow Ref.~\cite{Langer_2006}, which provides a function for the fraction of the star-formation rate density that has a metallicity less than $Z$ at given redshift, $z$:
\begin{equation}
   \Lambda (z, Z) =  \frac{\hat{\Gamma}[0.84, (Z/Z_{\odot})^2 10^{0.3 z}]}{\Gamma(0.84)},
\end{equation}
where $\hat{\Gamma}$ and $\Gamma$ are the incomplete and complete gamma functions, respectively.

\begin{figure*}
\centering
\includegraphics[height=8cm,width=19cm]{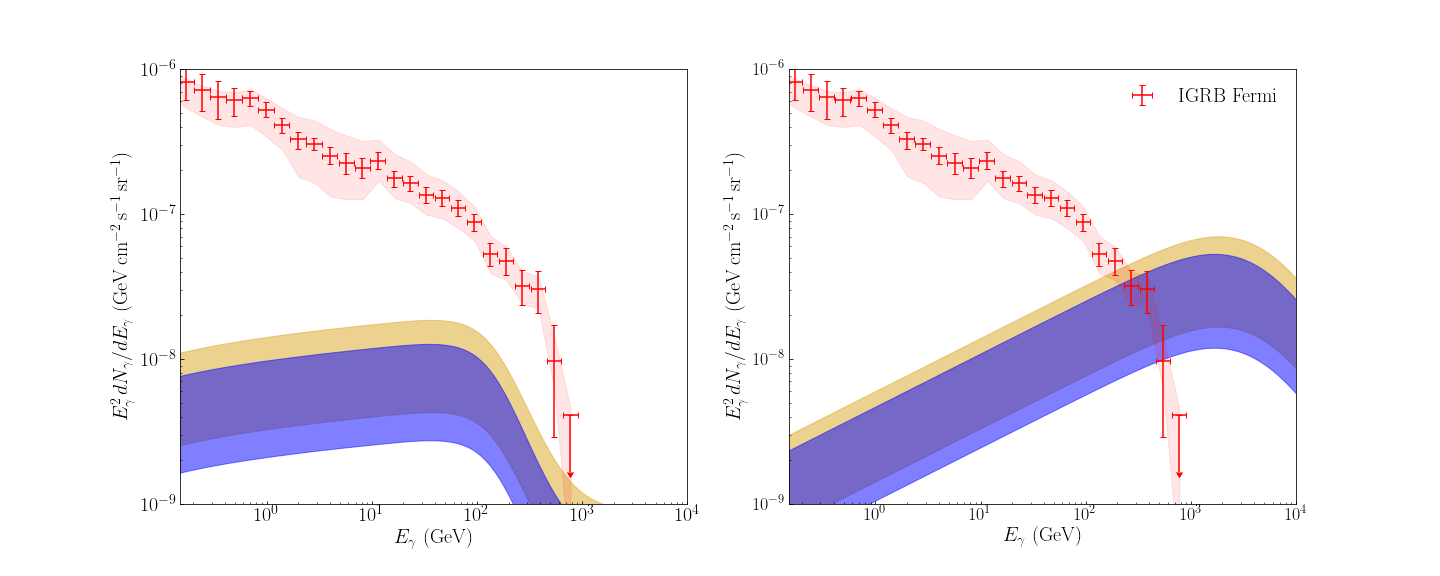}
\caption{The predicted contribution from TeV halos to the isotropic gamma-ray background, compared to the spectrum as measured and reported by the Fermi Collaboration~\cite{Ackermann_2015}. These results were derived using pulsar birth rates based on the measured star-formation rate, with (blue) and without (yellow) corrections for metallicity. The grey bands around the Fermi error bars represent the systematic uncertainty associated with the modelling of the Galactic foreground emission. In the left (right) frame, we show our results including (neglecting) the important effects of attenuation and electromagnetic cascades.}
\label{fig:TeVall}
\end{figure*}

Putting this all together, the final cosmic pulsar birth rate density is given by
\begin{equation}
   \frac{d\Gamma_p}{dV}(z) =  {\dot{\rho}_{*}(z)}\int \frac{d\Lambda}{dZ}(z, Z) \, f_{\rm NS}(Z) \,dZ. 
\end{equation}

In Fig.~\ref{fig:dNdV}, we plot the pulsar birth rate density as a function of redshift, using each of the three methods described in this section. The results are broadly consistent across these three methods, although the distribution based on the star-formation rate alone is somewhat larger at low redshifts than is found when using the metallicity corrected approach. The rate based on the rate of core collapse supernovae is consistent with both other methods, although with large uncertainties. In our main results, we will show the gamma-ray spectra predicted using cosmic pulsar birth rates as calculated using both the star-forming rate scaling, and the metallicity-corrected approach. 

In calculating the contribution from TeV halos to the isotropic gamma-ray background, it is necessary to take into account the effects of attenuation and the electromagnetic cascades that result from these interactions. In particular, TeV-scale photons can efficiently scatter with the infrared background to produce electron-positron pairs which then generate lower energy photons as they cool through the process of inverse Compton scattering. To account for this, we use the publicly available code $\gamma$-Cascade~\cite{Blanco_2019}, which fully models the effects of pair production, inverse Compton scattering, and synchrotron losses (see also, Refs.~\cite{Murase:2011yw,Murase:2012xs,2012ApJ...749...63M,Murase:2012df,Berezinsky:2016feh,Blanco:2017bgl}). This code adopts a background radiation field based on the model of Ref.~\cite{Dom_nguez_2010}, and adopts an extragalactic magnetic field of $10^{-13} \, {\rm G}$. In the case of the emission from TeV halos in M31 (as shown in Fig.~\ref{fig:M31 spec}), the effects of attenuation are negligible due to the proximity of this source. In contrast, in our calculation of the contribution to the IGRB from this class of sources, these interactions very substantially suppress the amount of emission that is predicted at energies above $\sim$100 GeV. In addition, the gamma rays that are produced through electromagnetic cascades significantly enhance the gamma-ray emission that is expected at lower energies. 

In the left frame of Fig.~\ref{fig:TeVall}, we show the main result of this paper, which is our estimate for the contribution from TeV halos to the IGRB. In the right frame of Fig.~\ref{fig:TeVall}, we show the same thing but, for comparison, neglecting the effects of attenuation and the subsequent contribution from electromagnetic cascades. While TeV halos produce very little of the emission that is observed by Fermi at low energies, this class of sources could be responsible for up to $\sim$20\% of the IGRB at 100 GeV, and perhaps even an larger fraction at TeV-scale energies.

\section{Implications for Millisecond Pulsar Populations}

Thus far, we have focused in this paper on the TeV halos associated with young and middle aged pulsars. In addition to these source classes, there exist pulsars with millisecond-scale periods which have obtained their angular momentum through interactions with a binary companion. Such ``recycled'' pulsars have lower magnetic fields, are much longer lived than their young and middle aged counterparts. 

Recent analyses of HAWC data have provisionally indicated that millisecond pulsars (MSPs) generate TeV halos with an efficiency and other characteristics that are similar to those associated with young and middle-aged pulsars~\cite{hooper2021evidence,Hooper:2018fih}. In our calculation of the contribution from TeV halos to the IGRB, we have not yet included any contribution from MSPs. If, however, the total spindown power of the MSP population is comparable to or larger than that associated with the young and middle aged pulsar population, these sources could significantly increase our estimate for the contribution of TeV halos to the IGRB.

The total spindown power of the Milky Way's MSP population is somewhat uncertain, in particular in regards to those pulsars located in the Inner Galaxy. Among the 283 MSPs in the ATNF catalog with a reported value of $\dot{E}$, the total spindown power is $2.2 \times 10^{37} \, {\rm erg}/{\rm s}$. Given the highly incomplete nature of this catalog, the total spindown power of all MSPs in the Milky Way is likely to be larger than this number by a factor of at least several, and perhaps significantly more. Comparing this to the total spindown power of the young and middle aged pulsars in the Milky Way, $\langle E_{{\rm rot},0}\rangle \, \Gamma_{p, {\rm MW}} \sim (7 \times 10^{48} \, {\rm erg}/{\rm s}) \, (1.4 \, {\rm century}^{-1}) \sim 3 \times 10^{39} \, {\rm erg}/{\rm s}$, we consider it plausible that MSPs could constitute a significant fraction of the total spindown power of the overall pulsar population. If it is robustly confirmed that MSPs generate TeV halos~\cite{hooper2021evidence,Hooper:2018fih}, this would lead us to potentially increase our estimate for the contribution of TeV halos to the IGRB (and from the TeV halo emission from M31). 

Further complicating this calculation is the fact that the MSP population density is not expected to scale with the current star-formation rate. Instead, the number of MSPs in a given galaxy will reflect the integrated star-formation history and the subsequent rate of stellar encounters within that environment (for example, see Ref.~\cite{Bahramian:2013ihw}).  

With these uncertainties acknowledged and in mind, we will proceed to estimate the gamma-ray emission from all TeV halos (including those associated with MSPs) by simply scaling our previous results by a factor that is equal to the total spindown power in all pulsars (including MSPs) divided by the total spindown power neglecting MSPs. Using the Milky Way pulsar populations to base this estimate, we note that the median MSP in the ATNF catalog is located only 3.6 kpc from Earth (considering only those MSPs with a reported distance measurement), corresponding to only the nearest 5\% of the Galactic Plane, and clearly indicating that most of the MSPs in the Milky Way have not yet been detected. With this in mind, we estimate that including MSPs would increase the fluxes shown in Figs.~\ref{fig:M31 spec} and~\ref{fig:TeVall} by a factor of roughly $\sim 1+ [(2.2 \times 10^{37})/0.05]/(3 \times 10^{39} \,f_{\rm beam}) \sim 1.5$, where $f_{\rm beam} \sim 0.3$ is the beaming fraction of a typical MSP.

\section{Implications for IceCube's Diffuse Neutrino Flux}

The results presented in the previous sections have potentially significant implications for the fields of high-energy gamma-ray and neutrino astrophysics. Studies utilizing observed correlations between gamma-ray and multi-wavelength emission have concluded that the IGRB is dominated by emission from a combination of star-forming galaxies and non-blazar active galactic nuclei (sometimes referred to as misaligned AGN). In particular, a recent study by Blanco and Linden concluded that star-forming galaxies produce $56^{+40}_{-23}\%$ of the IGRB at 10 GeV, while non-blazar AGN contribute $18^{+38}_{-12}\%$ of this signal at the same energy~\cite{blanco2021gammarays}. In contrast, the contributions to the IGRB from blazars (including both BL Lacs and flat-spectrum radio quasars)~\cite{Cuoco:2012yf,Harding:2012gk,Ajello_2012,Abdo_2010}, mergering galaxy clusters~\cite{Keshet:2002sw,Gabici:2002fg,Gabici:2003kr}, cosmic-ray interactions with circum-galactic gas~\cite{Feldmann_2012}, and ultra-high energy cosmic ray propagation~\cite{Taylor:2015rla,Ahlers_2011,Gelmini_2012} are each relatively small in comparison to these two source classes (see also, Refs.~\cite{Hooper:2016gjy,Linden:2016fdd,Tamborra:2014xia,DiMauro:2013xta,Inoue_2011}). 

The diffuse flux of high-energy astrophysical neutrinos reported by the IceCube Collaboration features an approximately power-law form over energies between tens of TeV and several PeV~\cite{IceCube:2015gsk,IceCube:2015qii,IceCube:2014stg,IceCube:2013low}, and exhibits flavor ratios that are consistent with the predictions of pion decay~\cite{IceCube:2015rro}. The lack of observed correlations in direction or time with known gamma-ray bursts~\cite{GRB2012} or blazars~\cite{Smith:2020oac,Hooper:2018wyk,Glusenkamp:2015jca} has strongly disfavored the possibility that many of these events originate from members of these source classes. This leaves star-forming galaxies and non-blazar AGN as the leading candidates for the origin of IceCube's diffuse high-energy neutrino flux. If any combination of these two source classes is responsible for generating the signal reported by IceCube, then these objects must also contribute significantly to the IGRB as measured by Fermi. More specifically, if cosmic-ray interactions in these sources produce pions in optically thin environments, the decaying pions will produce neutrinos, $\pi^+ \rightarrow \mu^+ \nu_{\mu} \rightarrow e^+ \nu_e \bar{\nu}_{e} \nu_{\mu}$, and gamma rays, $\pi^0 \rightarrow \gamma \gamma$, in a calculable ratio. Based on this relationship, quantitative studies have shown that if these source classes are responsible for IceCube's diffuse neutrino flux, they will also approximately saturate the IGRB, in particular at energies above several GeV (see, for example, Ref.~\cite{Hooper:2016jls}).

The results presented in this study indicate that TeV halos contribute significantly to the IGRB at the highest energies measured by Fermi. On similar grounds, TeV halos have previously been shown to dominate the diffuse TeV-scale emission observed along the Galactic Plane by the Milagro telescope~\cite{Linden:2017blp}. In this sense, it appears that TeV halos are a significant vector by which the process of star formation leads to the production of very high-energy gamma-ray radiation.

A critical point in this context is that TeV halos are leptonic sources, relying on inverse Compton scattering rather than pion production to generate their observed gamma-ray emission~\cite{sudoh2021highest,Hooper:2017gtd}. This forces us to conclude that a significant fraction of the highest energy gamma-ray emission observed from star-forming galaxies is not hadronic in origin, but is instead leptonic, suppressing the degree to which this class of sources could potentially contribute to IceCube's diffuse neutrino flux. By comparing the gamma-ray emission from star-forming galaxies~\cite{blanco2021gammarays} to that predicted in this study from the TeV halos, one can place an upper limit on the hadronic component of the emission from star-forming galaxies. Although the relevant uncertainties remain quite large at this time, this comparison is suggestive of a significantly leptonic origin of the TeV-scale emission from this class of sources. This conclusion would only be further strengthened if we were to include an estimated contribution from the TeV halos associated with MSPs. If future observations continue to support the conclusion that MSPs produce TeV halos, this could potentially disfavor star-forming galaxies as the primary source of IceCube's diffuse flux, and (by default) favor non-blazar AGN as the main sources of these mysterious particles.

\section{Summary and Conclusions}

In this study, we have used the observed characteristics of the Milky Way's TeV halos to estimate the gamma-ray emission from the population of these objects in the Andromeda Galaxy, as well as the contribution from TeV halos to the isotropic gamma-ray background (IGRB). In the case of Andromeda, we project that the Cherenkov Telescope Array (CTA) will be sensitive to the diffuse, multi-TeV emission from the TeV halos in that system. We also conclude that a significant fraction of the IGRB is generated by TeV halos. In particular, we estimate that at the highest energies measured by Fermi, $E_{\rm gamma} \sim 0.1-1 \, {\rm TeV}$, on the order of ${10\%}$ of the IGRB is generated by TeV halos. 

Taking this result into account would bring one to reduce their estimate for the neutrino flux from star-forming galaxies, potentially providing support for the hypothesis that misaligned AGN may be responsible for the diffuse neutrino flux reported by the IceCube Collaboration. If it is confirmed that millisecond pulsars also generate TeV halos, this would further increase the degree to which TeV halos are estimated to contribute to the IGRB.

\acknowledgments

We would like to thank Tim Linden and Rich Kron for helpful comments and discussions. DH is supported by the Fermi Research Alliance, LLC under Contract No. DE-AC02-07CH11359 with the U.S. Department of Energy, Office of High Energy Physics.

\bibliography{tevhalo}

\end{document}